\documentclass[aps,prl,preprint,groupedaddress,showkeys]{revtex4-1}

\usepackage{amsmath}
\usepackage[dvipdfmx]{graphicx}
\usepackage{url}
\usepackage{color}

\begin{document}


\title{Search for Anisotropy in the Ultra High Energy Cosmic Ray Spectrum using the Telescope Array Surface Detector}



\author{R.U.~Abbasi$^{1}$}
\author{M.~Abe$^{2}$}
\author{T.~Abu-Zayyad$^{1}$}
\author{M.~Allen$^{1}$}
\author{R.~Azuma$^{3}$}
\author{E.~Barcikowski$^{1}$}
\author{J.W.~Belz$^{1}$}
\author{D.R.~Bergman$^{1}$}
\author{S.A.~Blake$^{1}$}
\author{R.~Cady$^{1}$}
% ...
\affiliation{$^{1}$ High Energy Astrophysics Institute and Department of Physics and Astronomy, University of Utah, Salt Lake City, Utah, USA }
\affiliation{$^{2}$ The Graduate School of Science and Engineering, Saitama University, Saitama, Saitama, Japan }
\affiliation{$^{3}$ Graduate School of Science and Engineering, Tokyo Institute of Technology, Meguro, Tokyo, Japan }
% ...

\collaboration{The Telescope Array Collaboration }
\date{\today}
\begin{abstract}
The Telescope Array (TA) experiment is located in the western desert of Utah,
USA, and observes ultra high energy cosmic rays (UHECRs) in the Northern
hemisphere. At the highest energies, $E>10$~EeV, the shape of cosmic
ray energy spectrum may carry an imprint of the source density distribution
along the line of sight different in different directions of the sky. In
this study, we search for such directional variations in the shape of the
energy spectrum using events observed with the Telescope Array's surface
detector. We divide the TA field of view into two nearly equal-exposure
regions: the ``on-source'' region which we define as $\pm 30^\circ$ of the supergalactic
plane containing mostly nearby structures, and the complementary
``off-source'' region where the sources are further away on average. We compare
the UHECR spectra in these regions by fitting them to the broken power law and
comparing the resulting parameters. We find that the off-source spectrum has
an earlier break at highest energies. The chance probability to obtain such or
larger difference in statistically equivalent distributions is estimated as
$6.2\pm1.1\times10^{-4}$ ($3.2\sigma$) by a Monte-Carlo simulation. The observed
difference in spectra is in a reasonable quantitative agreement with a
simplified model that assumes that the UHECR sources trace the galaxy
distribution from the 2MRS catalogue, primary particles are protons and the
magnetic deflections can be neglected.
\end{abstract}

\pacs{96.50.sb}
\keywords{Ultra High Energy Cosmic Ray, Large Scale Structure, Anisotropy, Spectrum, Composition}

\maketitle
\section{Introduction}
\label{sec:introduction}
Ultra high energy cosmic ray (UHECR) primaries lose a notable fraction of energy in interactions with photons of cosmic microwave background radiation (CMBR) while propagating over distances comparable to the size of local cosmological structures 
such as voids and clusters of galaxies. 
The attenuation length depends on the particle type and energy. Protons which have energy $> 10^{19.7}$ eV lose the major part of their energy in pion photoproduction.
Consequently, the spectrum of protons is expected to show the suppression of
flux at these energies, which is known as the GZK cut off
\cite{1966PhRvL..16..748G,1966JETPL...4...78Z}.   Another relevant process for protons propagating in the CMBR is $e^{+}   e^{-}$ pair creation. 
This reaction is important for protons with $E \simeq10^{18.6}$ eV. 
Heavier nuclei also loose energy in
interactions with photon background fields through the photo-disintegration
processes \cite{PhysRev.180.1264} that typically lead to splitting off of 
individual nucleons. The mean free path of this process also becomes 
smaller at higher energy. 

The losses alter the UHECR energy spectrum in a way that depends on the distance
to the source. As a result, UHECR energy spectra may be different in different
areas on the celestial sphere: harder in the direction of nearby structures 
and softer where the large-scale concentrations of matter are further away.
In this work, we confront this expectation with the TA data by comparing energy
spectra of UHECR in regions which contain large number of nearby objects with
those corresponding to local voids. This approach is complementary to the
anisotropy studies focused on the distribution of arrival directions only.
\section{Experiment and dataset }
\label{sec:experiment-dataset}
Telescope Array (TA) experiment \cite{2041-8205-768-1-L1} employs a hybrid
approach to the detection of UHECR with energies E $> 10^{18}$ eV.  Cosmic
rays are observed using both fluorescence telescopes and a surface
detector array. The surface detector of TA consists of 507 scintillation counters
deployed on a square grid with 1.2~km spacing, covering an area about 670
km$^{2}$~\cite{AbuZayyad201287}. The operation of the surface detector 
started in 2008. Its duty cycle is 95\% on average. TA has accumulated
the largest exposure in the Northern hemisphere. The fluorescence telescopes
have a much smaller duty cycle of $\sim 9\%$ \cite{Abbasi201527}. 
In this analysis, cosmic-ray events with energies,  $E > 10^{19}$ eV observed by the surface detector of TA in the
period from May 2008 to May 2013 are used.

\begin{figure}
\includegraphics[angle=0,scale=0.8]{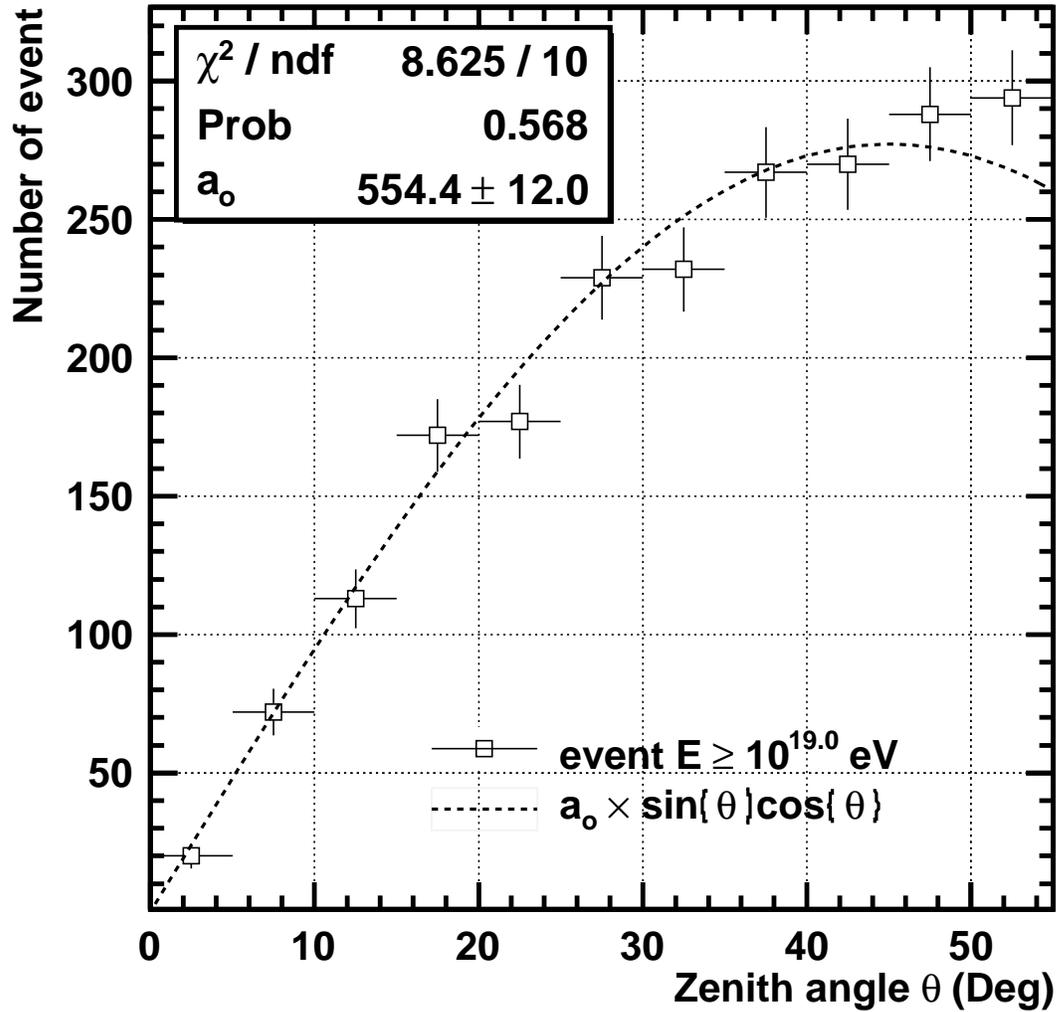}
\caption{The zenith angle distribution of observed shower events with energy E
  $\geq$ 10 EeV.  \label{1.epsL}}
\end{figure}

From Monte Carlo simulations, the trigger efficiency of cosmic-ray showers at
zenith angles of less than $55^{\circ}$ reaches 100\% in the energy range
greater than $10^{19}$~eV.  Corresponding estimated energy resolution is about
20\%, while the angular resolution is better than
2$^{\circ}$\cite{2041-8205-768-1-L1,2014arXiv1403.0644T}. The distribution of
the zenith angles of the observed events is shown in
Figure.~\ref{1.epsL}. It agrees well with the geometrical
exposure which is also shown on the plot for comparison.
\section{Analysis for Super Galactic Plane (SGP)}
\label{sec:analys-super-galact}
In this analysis, we divide the sky into two parts: the one containing
a larger number of nearby objects and another one with a lesser
number of nearby objects. These parts will be referred to as the ``{\bf on-source}'' and
``{\bf off-source}'' areas, respectively. We then compare the energy
spectra in the on-source and off-source sky regions.
As the division criterion we use the positions of sources with respect to the
Super Galactic Plane (SGP). The SGP is a major structure in the nearby
Universe containing a number of massive galaxy clusters at distances of a few
tens of Mpc \cite{1953AJ.....58...30D}. For the analysis, we choose the
on-source region as the region of the sky containing the SGP.  The exposure of the
TA experiment is almost equally divided when we define a sky within
$\pm$30$^{\circ}$ of the SGP as the on-source area and the rest as the
off-source area.  
The fractions of the total exposure corresponding to the on-
and off-source areas are 52\% and 48\%, respectively. Another --- technical
--- advantage of this choice is that the zenith angle distributions in the two
regions are practically identical. 
(Detailed discussion of systematic effects is given in the next section.)
In principle, one may base the choice of the on- and off-source regions
directly on the matter distribution in the nearby Universe. 
However, to use matter distribution directly  in the simple approach adopted here, all the details about the UHECR propagation, composition at high energies, etc, are needed. 
Given the existing uncertainties and limited statistics, we find the more simple approach adopted here to be more than adequate at the present stage.

\begin{figure}
\includegraphics[angle=0,scale=0.8]{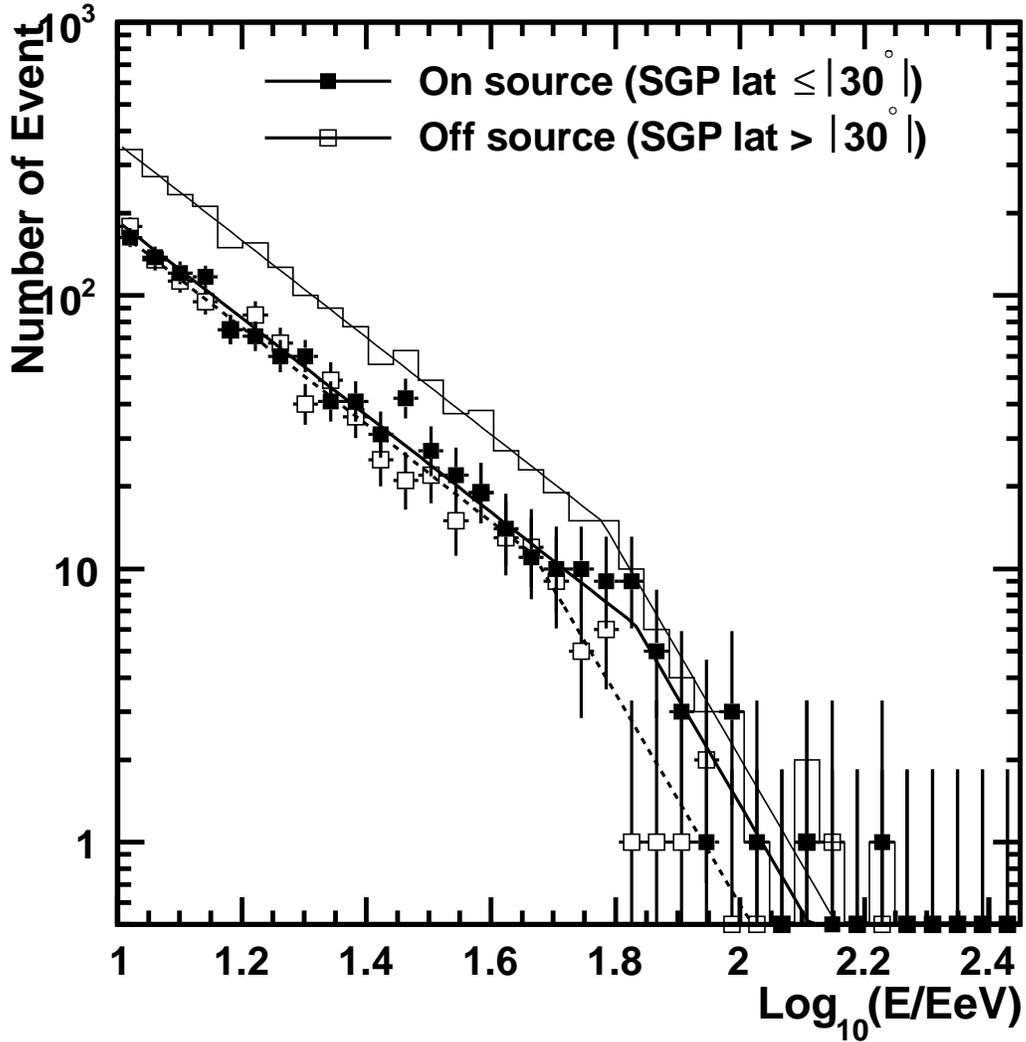}
 \caption{The energy distributions of observed shower events for the on- and
   off-source areas. The black histogram shows distribution of all
   events. Closed and open symbols show energy distributions observed in On-source
   and Off-source regions
   respectively. \label{2.epsL}}
\end{figure}

\begin{figure}
 \includegraphics[angle=0,scale=0.8]{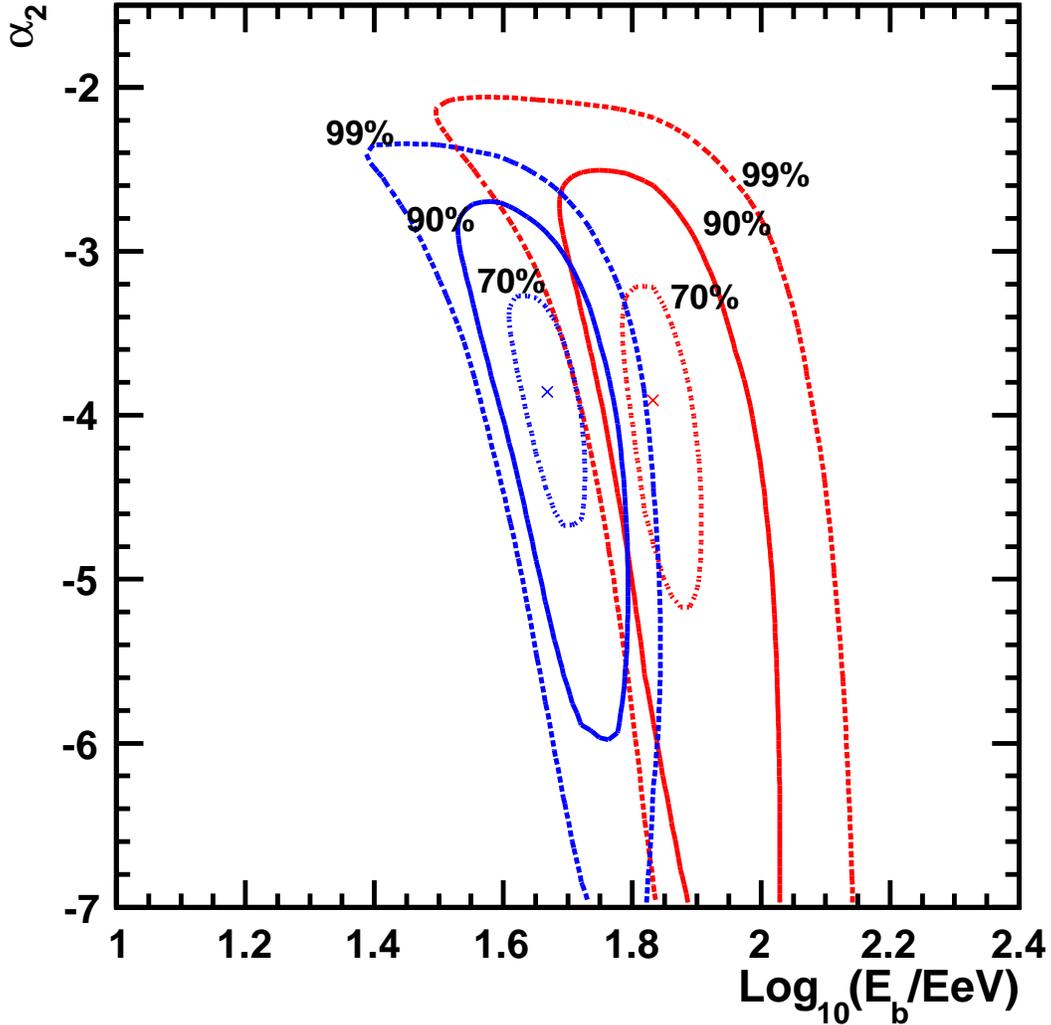}
 \caption{Contours of $\delta \log L$ in the plane of $E_{b}$ and $\alpha_{2}$
   in the SGP case.  Blue and red colors denote 70\%, 90\% and 99\% confidence
   levels for the off- and on-source regions,
   respectively.  \label{3.epsL}}
\end{figure}

Once the on- and off-source regions are fixed, we first compare the energy
distributions of the observed events coming from these regions.
Figure.~\ref{2.epsL} shows the energy
distribution of all the observed showers obtained for the entire exposure
(thin histogram), and separately for the On-source and Off-source areas (filled and
empty squares, respectively). 
The lines show the fits to these distributions
by a broken power law, with the fitting function defined as follows:
\begin{eqnarray}\label{BPLFUNC}
\begin{aligned}
  \frac{\Delta N(E)}{\Delta \log_{10}\left(\frac{E}{E_{0}}\right)}  
=  C_{0}\left( \theta(E_b-E) \left(\frac{E}{E_{0}}\right)^{-\alpha_{1}} +
\theta(E-E_b) \left(\frac{E_{b}}{E_{0}}\right)^{-\alpha_{1}+\alpha_{2}}\left(\frac{E}{E_{0}}\right)^{-\alpha_{2}}\right)
\end{aligned}
\end{eqnarray}
Here, $E_{0} = 1$~EeV, $C_{0}$ is a normalization constant proportional to the
total number of events, while $\alpha_{1,2}$ are spectral indexes below and
above the break, respectively.  The best fit parameters for the energy
distribution obtained for the entire exposure are
\begin{equation}
\label{eq:fit_params}
\begin{array}{rclcrcl}
C_{0}&=&2.14^{+0.34}_{-0.30}\times10^{4}& \qquad &
\alpha_{1}&=&-1.775^{+0.053}_{-0.053}\\
\log_{10}(E_{b}/{\rm EeV})&=& 1.778^{+0.040}_{-0.068}& \qquad &
\alpha_{2}&=&-3.91^{+0.64}_{-0.66}
\end{array}
\end{equation}
This fit is shown by the thin solid 
line on Figure.~\ref{2.epsL}. 
When fitting the energy distributions in the on- and off-source regions, the
slope before the break, $\alpha_{1}$, is set to the value obtained from the fit
to the distribution for the entire exposure, equation.~(\ref{eq:fit_params}). 
At higher energy in this energy range, mean free path of cosmic ray become shorter. It is expected that the differences at high energies due to different attenuation of flux for close and far sources, while negligible differences at low energies are expected. There is additionally the effect of increased isotropization in a coherent magnetic field at the lowest energy bins especially in case of nuclei other than proton. However, non-zero difference in spectrum slope would be exist. The small difference at lower energy side will be reflected to a change of $\alpha_{2}$ and $E_{b}$ those are free parameter in the fitting.
The normalization, $C_{0}$, is scaled to the corresponding fraction of the exposure in each region, while
$\log_{10}(E_{b}/E_{0})$ and $\alpha_{2}$ are set free and obtained from the
fitting in corresponding areas. 
There are bins with zero count at highest energy bins. Those bins are also included in the likelihood calculation.
The resulting broken power law functions are
plotted in Figure.~\ref{2.epsL} as solid and
dashed lines.
The results of the fit are summarized in
Table.~\ref{SGP_OnOffEnergyDistribution.tbl}.
\begin{table}
    \begin{tabular}{crrlr}
      \hline\hline
      Region & $C_{o}$ & $\alpha_{1}$ & $\log_{10}(E_{b}/EeV)$ & $\alpha_{2}$ \\
      \hline
      All           & $2.14^{+0.34}_{-0.30}\times10^{+4}$ & $-1.775^{+0.053}_{-0.053}$ & $1.778^{+0.040}_{-0.068}$  & $-3.91^{+0.64}_{-0.66}$ \\
      On source     & $(1.1128\times10^{+4})$           & $(-1.775)$             & $1.832^{+0.069}_{-0.041}$  & $-3.91^{+0.70}_{-1.30}$ \\
      Off source    & $(1.0286\times10^{+4})$           & $(-1.775)$             & $1.668^{+0.052}_{-0.053}$  & $-3.86^{+0.58}_{-0.82}$ \\
      \hline
    \end{tabular}
    \caption{Parameters of the best fit broken power law in the SGP case.}
    \label{SGP_OnOffEnergyDistribution.tbl}
\end{table}
Figure.~\ref{3.epsL} shows the best-fit
values of parameters together with the confidence contours in the
$\log_{10}(E_{b}/E_{0}) - \alpha_{2}$ plane.  
As one can see, there is a
difference in break energy between the on-source and off-source areas, $\Delta
\log_{10}(E_{b}/E_{0}) = 0.16$. Off-source, the break occurs at lower
energies, in agreement with expected larger attenuation for larger source
distances. The fraction of events above the break in the off-source region, $
N_{off}(E>E_{b}) /N_{all}(E> E_{b})$, is $0.337\pm0.050$ instead of $0.48$ as expected
from the exposure ratio.

To estimate the chance probability that the observed difference in the energy distribution 
occurred as a result of a fluctuation, we performed the following simulation.
In each energy bin the events have been randomly re-labeled as the on- and
off-source events following the binomial distribution with the parameters that
correspond to the ratio of the corresponding exposures, that is the on-source probability
0.52 and the off-source probability 0.48. After this re-shuffling, the new On-source
and Off-source energy distributions have been constructed and fitted by the broken power
laws in exactly the same way as the original data, giving the new values of
the break energies and the numbers of events after the break. This 
procedure then has been repeated $5\times 10^4$ times. 
\begin{figure}
\includegraphics[angle=0,scale=0.8]{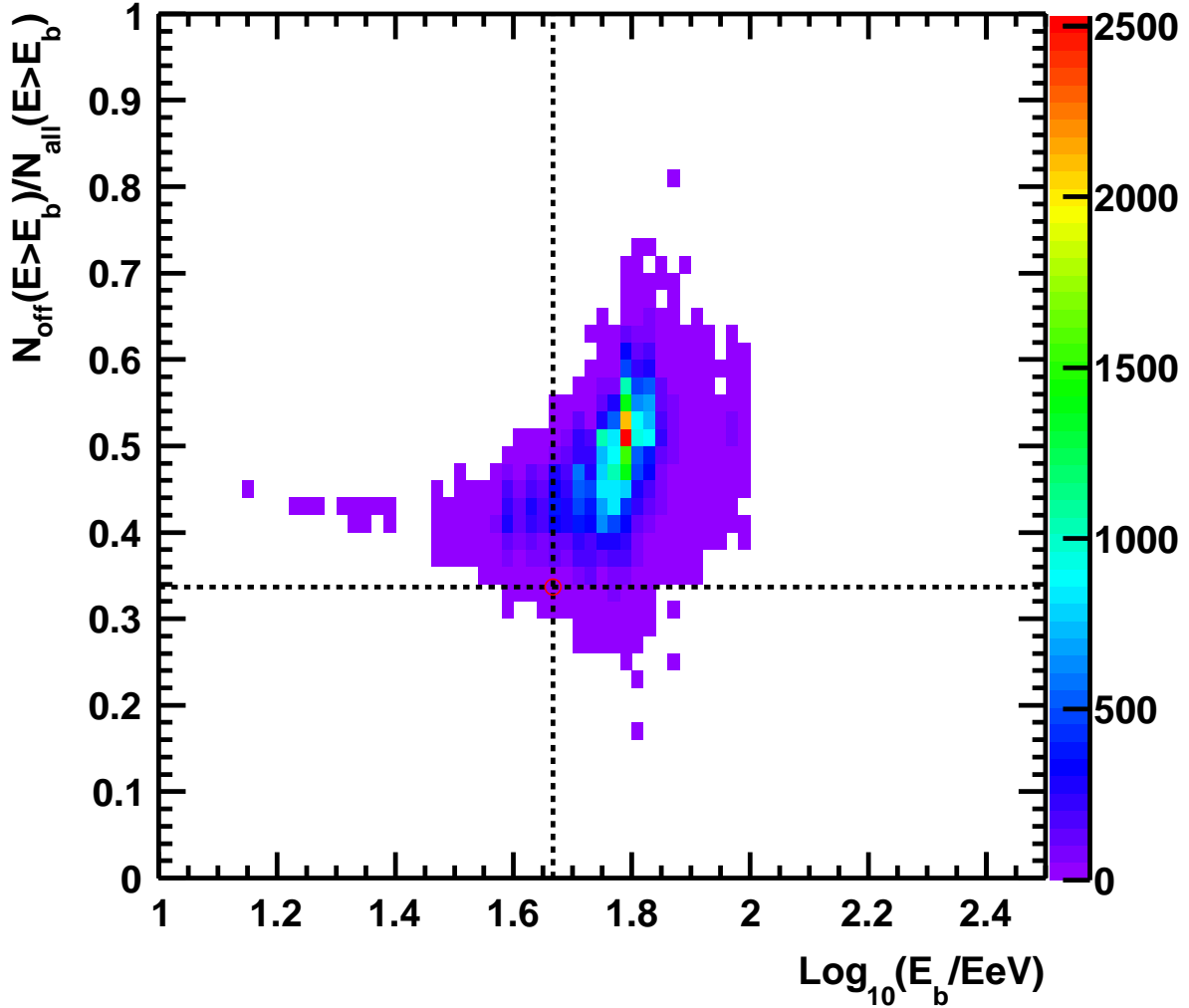}
\caption{Event fraction in the off-source region above the break energy
  versus $\log_{10}(E_{b}/EeV)$ obtained in Monte Carlo simulations. Red point and dashed line represents observed value in real data.}
\label{4.epsL}
\end{figure}
Figure.~\ref{4.epsL} shows the distribution  of occurrences of
parameters that characterize the off-source energy distribution shape --- the 
break energy $E_b$ and the fraction of events above the break 
$N_{off}(E>E_{b}) /N_{all}(E> E_{b})$. The horizontal axis corresponds to the
off-source break energy and vertical axis to the event fraction
above the break energy. The values observed for the data are marked by the
horizontal and vertical dashed lines which divide the parameter space into
four regions.  

The number of occurrences of parameters in the resulting four regions of the
parameter space is summarized in Table~\ref{SGP_ChanceProbability.tbl}.
\begin{table}
 \begin{tabular}{c|r|c}
      \hline\hline
      Condition  & case  & Fraction  \\
      \hline
      {\scriptsize $E_{b}>10^{1.668}EeV$, $ \frac{N_{off}(E>E_{b})}{N_{all}(E> E_{b}} >{0.337}$}  & 45031  & $0.9008(\pm 0.0013)$   \\
      {\scriptsize $E_{b}<10^{1.668}EeV$, $ \frac{N_{off}(E>E_{b})}{N_{all}(E> E_{b}} >{0.337}$}  &  4606  & $0.0921(\pm 0.0013)$   \\
      {\scriptsize $E_{b}<10^{1.668}EeV$, $ \frac{N_{off}(E>E_{b})}{N_{all}(E> E_{b}} <{0.337}$}  &    31  & $0.00062(\pm 0.00011)$   \\
      {\scriptsize $E_{b}>10^{1.668}EeV$, $ \frac{N_{off}(E>E_{b})}{N_{all}(E> E_{b}} <{0.337}$}  &   352  & $0.00704(\pm 0.00037)$  \\
      \hline
\end{tabular}
    \caption{The number of occurrences $N$ with given break energy and number
      of events above the break, and the corresponding fractions. }
    \label{SGP_ChanceProbability.tbl}
\end{table}

Following the predictions of the UHECR propagation models that suggest that
the spectrum in the off-source region should have a lower break energy and a
smaller number of events above the break, we consider a trial successful if it
has both of these parameters smaller than in the data. The fraction of
successful trials (third line in Table~\ref{SGP_ChanceProbability.tbl}) in our
MC simulation is 31, which gives the p-value $p =
6.2\times10^{-4}$ (3.2$\sigma$).
\section{Systematic errors}\label{sec:systematic-errors}
\begin{figure}
\includegraphics[angle=0,scale=0.8]{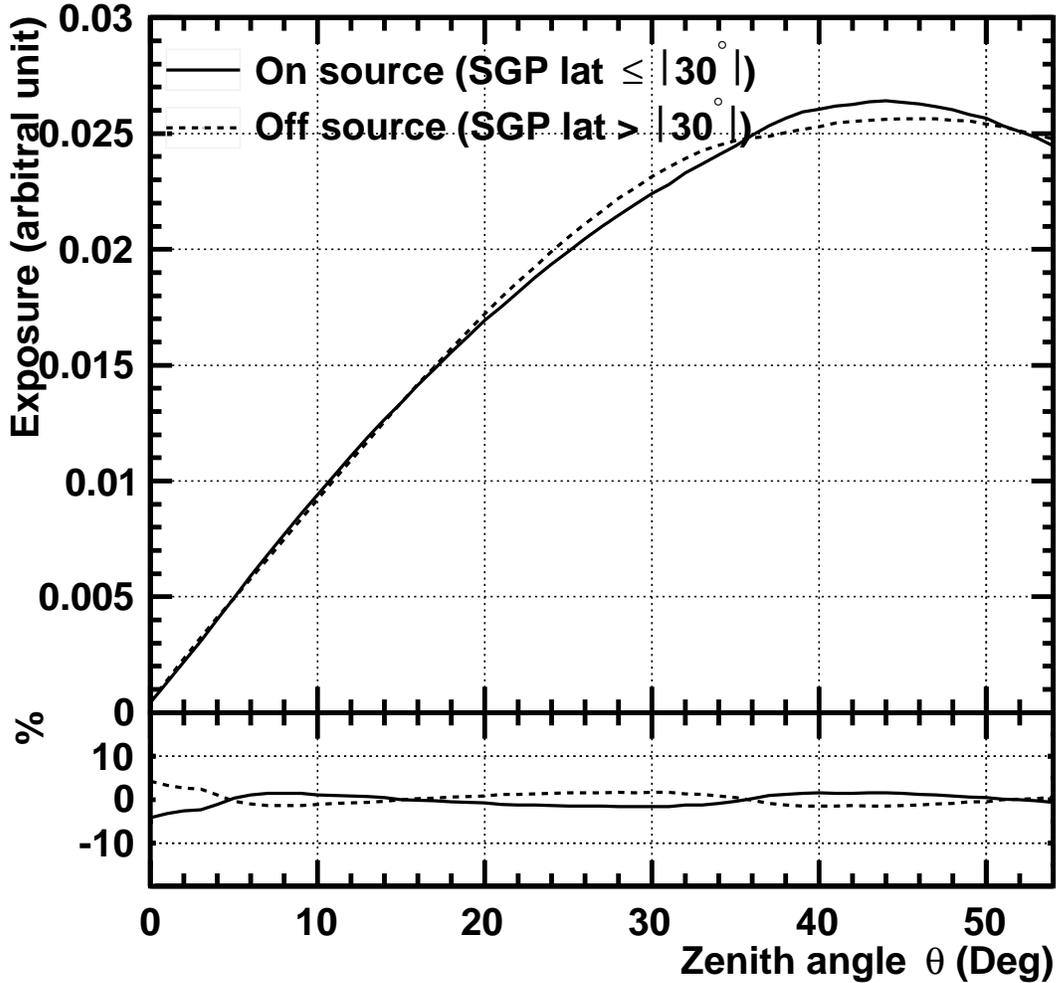}
\caption{The zenith angle distributions in On and Off source
 areas. The relative deviations of these
distributions from the total (geometrical) exposure are shown on the bottom
panel.\label{5.epsL}}
\end{figure}

As the spectrum of UHECR is potentially sensitive to reconstruction biases, an
important question is whether the on- and off-source regions correspond to the
same observational conditions, notably the same zenith angle distribution of
the events.  For the adopted on- and off-source regions, the distributions of
zenith angles of exposure are plotted in
Figure.~\ref{5.epsL}. 
The relative deviations of these
distributions from the total (geometrical) exposure are shown on the bottom
panel; one can see that they do not exceed several percent. The difference
between the two regions in observing condition is thus negligible considering
our statistics.

The time variation of the energy scale due to change in atmospheric conditions
can be another source of a systematic error. To check its
influence, the event rate with energies greater than $10^{19.0}$~eV was
studied in anti-sidereal time \cite{0370-1298-67-11-306}.  The amplitude of
fluctuations of the event rate in the anti-sidereal time was found to be at
most 5\% $\pm$3\%.  Given the observed spectral index around $10^{19.0}$ eV,
this corresponds to the energy shift by $\sim 2.5$\%.
One may estimate the effect of this possible energy shift by re-doing the
calculation of the p-value with the event energies shifted by $2.5$\%
upwards. The resulting p-value is $6.9\times10^{-4}$ (3.2 $\sigma$), i.e., the
effect of this uncertainty is negligible.

\section{Discussion}

\begin{figure}
    \includegraphics[angle=0,scale=0.8]{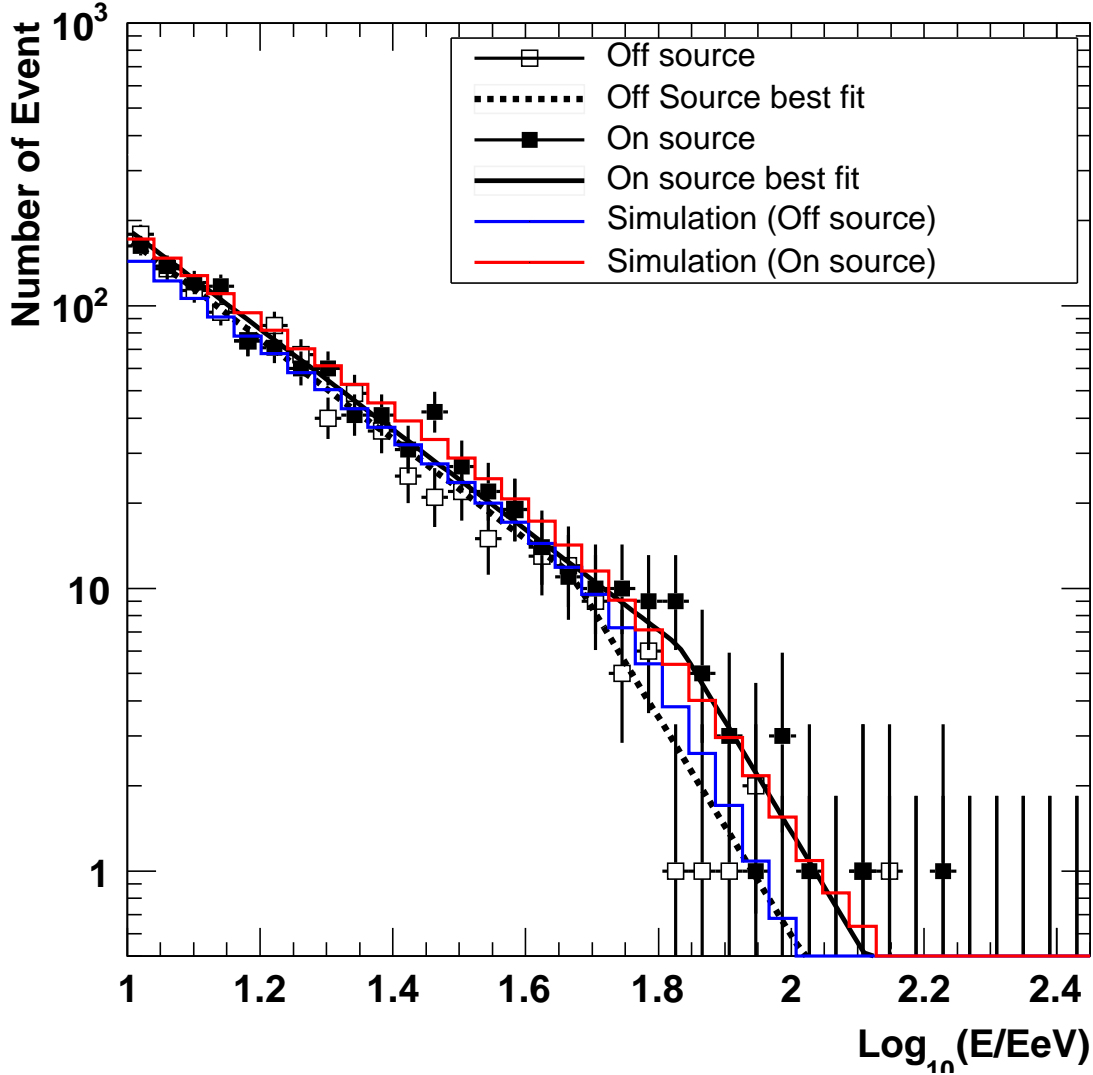}
    \caption{Comparison of energy distributions expected for protons simulated with CRPropa 2.2.0.4 arriving
      from the sources with injection index of -2.2, evolution parameter of 7 and 2MRS density profile.}
    \label{Data_2MRSComparisonP22M7.epsL}
\end{figure}

To see if the observed differences in energy distributions are compatible with model predictions , we performed a simplified numerical simulation using a propagation code CRPropa 2.2.0.4 \cite{Kampert201341} and the source distribution from the 2MRS catalogue \cite{0067-0049-199-2-26} using the density profile calculation described in \cite{Koers21102009}.
In a simplified expectation, when the composition at origin consists of nuclei, the expected difference of spectrum between on and off-source region start develop from lower energy and show gradual development as compared to the case of proton. That is because nuclei have photo-disintegration as a dominant energy loss process while proton have pion production with energy threshold.
Figure.~\ref{Data_2MRSComparisonP22M7.epsL}  displays the results. Here protons were assumed as primary particles. 
The injection index and evolution parameter were set to $-2.2$ and $7$, respectively, as to reproduce the observed TA energy spectrum \cite{EKido201410UHECR2014}. 
Simulations are done in one dimension assuming 0.1 nG random magnetic field.
The difference of the observed energy distributions in the on-source and
off-source regions were, qualitatively,  well reproduced  by this simulation,
given the simplifications made when modelling the propagation.  
A quantitative comparison with the theoretical predictions would require a
more accurate modeling of the propagation, as well as a better
understanding of the UHECR composition not available at present TA statistics. 

\section{Summary}

Using data obtained by the TA surface detector in first 5 years, a new approach to search for the anisotropy of UHECR is developed. 
It employs the modulation of the energy spectrum due to energy
losses in the interactions with the CMBR during propagation of primaries. Such
modulations occur in a different way for sources at different distances; as a
result, the UHECR spectra may differ in region of the sky that contain nearby
matter concentrations (sources are closer on average) and regions that do not
contain such structures (sources are further on average).
The TA field of view was divided into two almost equal-exposure parts: the
on-source region within $\pm 30^{\circ}$ from the SGP, and the complementary
off-source region.  The energy distributions of the observed TA events in
these regions were fitted by a broken power law.  The results of the fit are
summarized in Table.~\ref{SGP_OnOffEnergyDistribution.tbl}. The parameters
that characterize the energy spectra in the two regions differ: in the
off-source region the flux has earlier suppression as compared to the
on-source region, in qualitative agreement with expectation from the
propagation models. The chance probability that this difference arises as a
result of a fluctuation in two statistically equivalent distributions was
estimated to be $\sim6.2\times10^{-4}$ (3.2$\sigma$).
We conclude that there is a $\sim 3\sigma$ indication of the spectrum differences of UHECR in different regions of the sky in the Northern hemisphere. 
In on-source region, it is known there is a direction with excess of number of event which is found in a paper \cite{2041-8205-790-2-L21}. 
The excess is evaluated using events with energy above 57 EeV and by integrating number of events with area of $20^{\circ}$. The nature of this excess is not known yet.
In this paper, the approach applied does not use specific threshold in energy and angular size. 
We believe that the approach proposed here can be developed further, and may help to reveal the sources of cosmic rays and their chemical composition.

\begin{acknowledgments}
\section{acknowledgments}
The Telescope Array experiment is supported by the Japan Society for the Promotion of Science through Grants-in-Aid for Scientific Research on Specially Promoted Research (21000002) ``Extreme Phenomena in the Universe Explored by Highest Energy Cosmic Rays'' and for Scientific Research (19104006), and the Inter-University Research Program of the Institute for Cosmic Ray Research; by the U.S. National Science Foundation awards PHY-0307098, PHY-0601915, PHY-0649681, PHY-0703893, PHY-0758342, PHY-0848320, PHY-1069280, PHY-1069286, PHY-1404495 and PHY-1404502; by the National Research Foundation of Korea (2015R1A2A1A01006870, 2015R1A2A1A15055344, 2016R1A5A1013277,  2007-0093860, 2016R1A2B4014967); by the Russian Academy of Sciences, RFBR grant 16-02-00962a (INR), IISN project No. 4.4502.16, and Belgian Science Policy under IUAP VII/37 (ULB). The foundations of Dr. Ezekiel R. and Edna Wattis Dumke, Willard L. Eccles, and George S. and Dolores Dor\'e Eccles all helped with generous donations. The State of Utah supported the project through its Economic Development Board, and the University of Utah through the Office of the Vice President for Research. The experimental site became available through the cooperation of the Utah School and Institutional Trust Lands Administration (SITLA), U.S. Bureau of Land Management (BLM), and the U.S. Air Force. We appreciate the assistance of the State of Utah and Fillmore offices of the BLM in crafting the Plan of Development for the site.  We also wish to thank the people and the officials of  Millard County, Utah for their steadfast and warm support. We gratefully acknowledge the contributions from the technical staffs of  our home institutions. An allocation of computer time from the Center for High Performance Computing at the University of Utah is gratefully acknowledged.
\end{acknowledgments}
\bibliography{spectrumaniso2015}
\bibliographystyle{apsrev4-1}

\end{document}